# The Immediate, Exemplary, and Fleeting Echelle Spectroscopy of SN 2023ixf: Monitoring Acceleration of Slow Progenitor Circumstellar Material Driven by Shock Interaction

Danielle Dickinson[1], Dan Milisavljevic[1,2], Braden Garretson[1], Luc Dessart[3], Raffaella Margutti[4], Ryan Chornock[4], Bhagya Subrayan[5], D. John Hillier[6], Eli Golub[7], Dan Li[7], Sarah E. Logsdon[7], Jayadev Rajagopal[7], Susan Ridgway[7], Nathan Smith[5], and Chuck Cynamon[8]

[1] Department of Physics and Astronomy, Purdue University, 525 Northwestern Ave, West Lafayette, IN 47907, USA
[2] Integrative Data Science Institute, Purdue University, West Lafayette, IN 47907, USA
[3] Institut d'Astrophysique de Paris, CNRS-Sorbonne Université, 98 bis boulevard Arago, F-75014 Paris, France
[4] Department of Astronomy, University of California, Berkeley, CA 94720-3411, USA
[5] Steward Observatory, University of Arizona, 933 North Cherry Avenue, Tucson, AZ 85721-0065, USA
[6] Department of Physics and Astronomy & Pittsburgh Particle Physics, Astrophysics, and Cosmology Center (PITT PACC), University of Pittsburgh, 3941 O'Hara Street, Pittsburgh, PA 15260, USA
[7] NSF NOIRLab, 950 N Cherry Ave, Tucson, AZ 85719, USA
[8] Supra Solem Observatory, Bradley, CA 93426, USA

## Abstract

We present high-resolution WIYN/NEID echelle spectroscopy ($R \approx 70{,}000$) of the supernova (SN) 2023ixf in M101, obtained 1.51 to 18.51 days after explosion over nine epochs. Daily monitoring for the first 4 days after explosion shows narrow emission features ($\lesssim 200\ \mathrm{km\ s^{-1}}$), exhibiting predominantly blueshifted velocities that rapidly weaken, broaden, and vanish in a manner consistent with radiative acceleration and the SN shock eventually overrunning or enveloping the full extent of the dense circumstellar medium (CSM). The most rapid evolution is in the He I emission, which is visible on day 1.51 but disappears by day 2.62. We measure the maximum pre-SN speed of He I to be $25^{+0}_{-5} \pm 2\ \mathrm{km\ s^{-1}}$, where the error is attributable to the uncertainty in how much the He I had already been radiatively accelerated and to measurement of the emission-line profile. The radiative acceleration of CSM is likely driven by the shock–CSM interaction, and the CSM is accelerated to $\gtrsim 200\ \mathrm{km\ s^{-1}}$ before being completely swept up by the SN shock to $\sim 2000\ \mathrm{km\ s^{-1}}$. We compare the observed spectra with spherically symmetric r1w6b `HERACLES/CMFGEN` model spectra and find the line evolution to generally be consistent with radiative acceleration, optical depth effects, and evolving ionization state. The progenitor of SN 2023ixf underwent an enhanced mass-loss phase $\gtrsim 4$ yr prior to core collapse, creating a dense, asymmetric CSM region extending out to approximately $r_{\mathrm{CSM}} = 3.7 \times 10^{14}\ (v_{\mathrm{shock}}/9500\ \mathrm{km\ s^{-1}})$ cm.

## 1. Introduction

Observations made shortly ($\lesssim 10$ days) after a core-collapse supernova (CCSN) explosion provide opportunities to probe the mass loss of the progenitor star during poorly understood terminal phases of evolution ($\lesssim 10^2$ yr; N. Smith 2014; T. J. Moriya et al. 2017). Narrow ($\lesssim 200\ \mathrm{km\ s^{-1}}$) emission lines seen in "flash" spectroscopy at this time, produced in slow-moving circumstellar material (CSM) that is excited by either shock-breakout ionization (D. Khazov et al. 2016) or photoionization of the CSM by shock–CSM interaction (C. Fransson & R. A. Chevalier 1989; N. N. Chugai & I. J. Danziger 1994; N. Smith et al. 2015), are particularly informative. These short-lived narrow features may include H I, He I, He II, C III, N III, O III, C IV, N IV, O V, and O VI. These lines are reminiscent of the persistent lines observed in Type IIn supernovae (SNe IIn; E. M. Schlegel 1990; N. Smith 2017), whose progenitors undergo an enhanced mass-loss phase ($\dot{M} \approx 0.01-1 M_\odot\ \mathrm{yr^{-1}}$, $v_{\mathrm{wind}} = 100\ \mathrm{km\ s^{-1}}$, $\rho = 10^{-14}$–$10^{-12}\ \mathrm{g\ cm^{-3}}$) in 10–100 yr preceding stellar death. Analysis of these lines can constrain the mechanisms responsible for these mass-loss events (C. Fransson et al. 2014; J. H. Groh 2014; L. Dessart et al. 2015; I. Shivvers et al. 2015; N. Smith et al. 2015), which in turn can provide insights into the nature of the progenitor star and its internal structure at the time of explosion (N. Smith & W. D. Arnett 2014).

The majority of CCSNe do not produce narrow lines throughout much of their evolution, but early-time narrow emission is observed in a third of the population (R. J. Bruch et al. 2021), of which SNe IIn comprise 4%–9% of the total core-collapse population (N. Smith 2014; I. Shivvers et al. 2017; D. A. Perley et al. 2020). Examples of well-studied CCSNe observed to have short-lived narrow emission include SN 1983K (V. S. Niemela et al. 1985), SN 1993J (S. Benetti et al. 1994), SN 1998S (D. C. Leonard et al. 2000), SN 2006bp (R. M. Quimby et al. 2007), SN 2013cu (A. Gal-Yam et al. 2014), PTF11iqb (N. Smith et al. 2015), SN 2013fs (O. Yaron et al. 2017; C. Bullivant et al. 2018), SN 2016bkv (G. Hosseinzadeh et al. 2018), SN 2020pni (G. Terreran et al. 2022), SN 2020tlf (W. V. Jacobson-Galán et al. 2022), SN 2024ggi

(W. V. Jacobson-Galán et al. 2024a; M. Shrestha et al. 2024), and SN 2024bch (L. Tartaglia et al. 2024; J. E. Andrews et al. 2025). The timescale over which this interaction is observed suggests that, like SNe IIn, this enhanced mass loss occurs up until stellar death. Unlike SNe IIn, however, the mass loss occurs on much shorter timescales ($<10$ yr), concentrating high-density material in a region close to the explosion (D. Khazov et al. 2016). Consequently, the associated narrow emission-line features disappear within a few days.

SN 2023ixf in the nearby galaxy M101 ($D \sim 6.9$ Mpc; A. G. Riess et al. 2022) was the brightest SN since SN 2014J (A. Goobar et al. 2014). SN 2023ixf was discovered on 2023 May 19 (UT dates are used throughout this paper) with an unfiltered magnitude of 14.9 mag (K. Itagaki 2023). Two hours later, D. A. Perley et al. (2023) classified SN 2023ixf as a Type II SN with flash ionization features. This early detection and classification enabled unparalleled rapid, high-cadence, multiwavelength monitoring of the rise to maximum light and subsequent decline. Nightly and even subnightly photometric and spectroscopic observations were obtained.

By comparing a time series of moderate-resolution spectroscopy to the HERACLES/CMFGEN model r1w6b, W. V. Jacobson-Galán et al. (2023) concluded that the progenitor star of SN 2023ixf underwent a 3–6 yr enhanced mass-loss (superwind) phase prior to explosion, at a rate of $\dot{M} \approx 10^{-2}\, M_\odot\, \mathrm{yr}^{-1}$, adopting a 50 km s$^{-1}$ wind speed. Narrow emission-line features disappeared $\sim$4 days following explosion, suggesting a CSM radius of $<10^{15}$ cm. R. S. Teja et al. (2023) and K. A. Bostroem et al. (2024), utilizing photometric and spectral models of the SN, also estimated enhanced mass-loss rates ranging between $10^{-2}\, M_\odot\, \mathrm{yr}^{-1}$ to $10^{-3}\, M_\odot\, \mathrm{yr}^{-1}$. E. A. Zimmerman et al. (2024) suggests a similar mass-loss rate of $10^{-1.96}\, M_\odot\, \mathrm{yr}^{-1}$ from measuring the centroid shift of several narrow lines. Broadly, these studies find that the mass lost immediately preceding SN 2023ixf was ejected with a rate far exceeding the typical red supergiant (RSG) mass-loss rate ($\leqslant 10^{-6}\, M_\odot\, \mathrm{yr}^{-1}$; E. R. Beasor et al. 2023).

N. Smith et al. (2023, hereafter S23), presented high-resolution echelle spectra ($R \sim 50{,}000$, covering 4300–9000 Å) obtained with the Potsdam Echelle Polarimetric and Spectroscopic Instrument (PEPSI) mounted on the Large Binocular Telescope (LBT) and starting 2.62 days following explosion. A pre-shock CSM speed of 115 km s$^{-1}$ was measured using the observed profile of the H$\alpha$ emission line. S23 suggests this speed likely arises from a combination of radiative acceleration and the original pre-SN CSM dynamics. The post-shock CSM speed of 1300 km s$^{-1}$ was measured from the bluest edge of the intermediate-width P Cygni profile. Due to systematic blueshifting in the line profiles and a lack of a narrow P Cygni profile in the narrow component, S23 proposed that the CSM is asymmetric and toroidal, and that the bulk of the CSM is not along our line of sight (LOS). Significant asymmetry in the CSM was confirmed by observations of a high early-time polarization (S. S. Vasylyev et al. 2023; M. Shrestha et al. 2025; A. Singh et al. 2024).

Archival pre-explosion images reveal a dusty candidate RSG progenitor (C. D. Kilpatrick et al. 2023; J. L. Pledger & M. M. Shara 2023; M. D. Soraisam et al. 2023). S. D. Van Dyk et al. (2024) present an extensive multiwavelength study and find the progenitor mass to be 12–15$M_\odot$ and a metallicity that is likely solar. From single-star models of the spectral energy distribution (SED), C. D. Kilpatrick et al. (2023), J. E. Jencson et al. (2023), and Y.-J. Qin et al. (2024) find progenitor masses of $11 \pm 2 M_\odot$, $17 \pm 4 M_\odot$, and $18^{+0.7}_{-1.2} M_\odot$, respectively. C. D. Kilpatrick et al. (2023) and J. M. M. Neustadt et al. (2024) determine mass-loss rates of $10^{-6}$ and $10^{-5} M_\odot\, \mathrm{yr}^{-1}$, respectively, from their SED modeling. Both suggest the progenitor underwent radiatively driven mass loss that appears to be consistent with more luminous and cooler RSGs late in their evolution. The archival data may predate the estimated window of the enhanced mass-loss phase, hence these derived parameters may trace CSM at larger radii than the dense CSM associated with the transient narrow emission features.

The RSG progenitor exhibited variability between 0.2 and 0.4 mag in the near-IR, with a period of $1091 \pm 71$ days (M. D. Soraisam et al. 2023). However, no such variability or changes consistent with precursor eruptions were detected across the electromagnetic spectrum in the 12 yr preceding collapse (N. Flinner et al. 2023; M. D. Soraisam et al. 2023; J. M. M. Neustadt et al. 2024; C. L. Ransome et al. 2024), despite the enhanced mass loss suspected to have occurred during this time frame.

In this paper, we present high-resolution echelle spectroscopy of SN 2023ixf beginning 1.51 days following explosion that provides the most precise estimate of the progenitor star's pre-SN CSM velocity and closely follows the radiative acceleration of the CSM. Observations are presented in Section 2. Analysis and results, including measurements of rapidly changing emission-line widths and estimates of radiative acceleration, are made in Section 3. We discuss our findings and interpret them in context with HERACLES/CMFGEN models and the Type IIn SN 1998S in Section 4. Our conclusions are provided in Section 5. We adopt a redshift of $z = 0.000869644 \pm 0.000042$, measured from narrow Na I D absorption associated with the host (see Section 2.1). We adopt MJD = 60082.75 (G. Hosseinzadeh et al. 2023) as the time of explosion ($t_{\mathrm{exp}} = 0$) and assume a distance of 6.9 Mpc to the host M101 (A. G. Riess et al. 2022; W. V. Jacobson-Galán et al. 2023; E. A. Zimmerman et al. 2024).

## 2. Observations

### 2.1. Spectroscopy

We obtained nine spectra of SN 2023ixf spanning 4000–9000 Å within the first 18 days of explosion from the NEID instrument (C. Schwab et al. 2016) mounted on the WIYN 3.5 m telescope on Kitt Peak, AZ.[9] Table 1 summarizes our observations. NEID is a thermo-mechanically stabilized spectrograph with a 9 k × 9 k CCD (10 $\mu$m pixels) designed to provide extreme-precision radial velocity measurements (G. Stefansson et al. 2016; P. Robertson et al. 2019). The NEID spectra were obtained using the high-efficiency mode ($R \sim 70{,}000$), which has a fiber aperture of 1″.5 on the sky.

The spectra were reduced with the NEID Data Reduction Pipeline, and the Level 2, 1D extracted spectra were used.[10] In order to remove the blaze from the echelle observations, we normalized the spectra and blaze with their respective Frobenius norm. We then divided the normalized blaze from the normalized spectrum. For all spectra shown, we divided the spectra by the mean-normalized telluric model provided with the NEID data.

[9] The WIYN Observatory is a joint facility of the NSF's National Optical-Infrared Astronomy Research Laboratory, Indiana University, the University of Wisconsin-Madison, Pennsylvania State University, Purdue University, and Princeton University.

[10] See https://neid.ipac.caltech.edu/docs/NEID-DRP/ for more information.

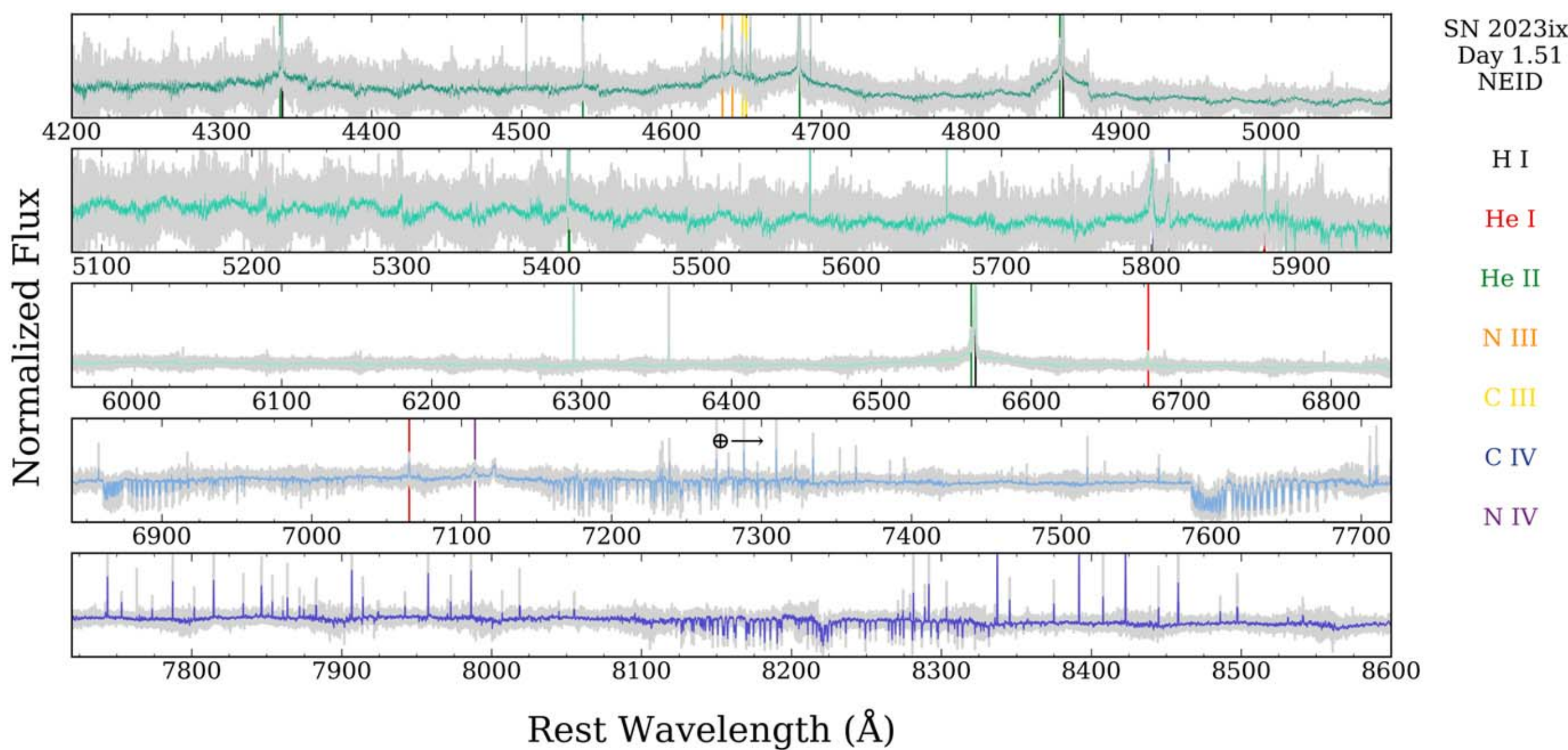

**Figure 1.** NEID 1.51 day spectrum with all orders stitched together. All other epochs listed in Table 1 have similar quality. The gray spectrum is the full resolution, and the overlaid color spectrum is convolved with a 1.66 Å filter. Flash ionization features are indicated with vertical lines.

The spectrum on day 1.51 is shown in the left panel of Figure 1. The gray spectrum is the full-resolution data, and the overlaid color has been convolved with a 1.66 Å filter. The apparent variability in the continuum of the spectrum is an artifact arising from stitching the orders of the spectrum together for this representation. All of our analysis in this work is done on individual orders and small wavelength ranges ($\pm1000\ \mathrm{km\ s^{-1}}$ spanning a given emission line), where the boundaries of each order, including the decreased sensitivity near the edges, do not contaminate the appearance of narrow features. We convolved the spectra with smoothing filters to best inspect the spectral features. The box sizes of the filters were chosen to what looked best for the purposes of displaying the data. All box sizes are below 0.5 Å and were chosen to be as small as possible. When preparing these figures, each spectrum was inspected at various smoothing box sizes (1–10 px) to validate that the convolution did not artificially create features.

A list of lines identified and their narrow-component measurements are presented on Table 2. The line wavelengths are from the NIST Atomic Spectral Database.[11] A $\sim1800\ \mathrm{km\ s^{-1}}$ Gaussian was subtracted from each of the lines to isolate the narrowest emission. For the lines denoted with an asterisk, we used `numpy.polyfit` to fit a linear model to the continuum for $200 < |v| < 1000\ \mathrm{km\ s^{-1}}$ instead. The values in the table are the FWHM of a Lorentzian or Gaussian profile centered at $0\ \mathrm{km\ s^{-1}}$. These functions were compared to the blue wing of each line profile, and the FWHM found by visual inspection. Whether a Lorentzian or Gaussian function was used was determined by visual goodness of fit. See Section 4 for discussion and justification for isolating the blue wing in this width measurement. All functional comparisons to spectra are shown in Appendix Figures 11, 12, and 13. The Pickering lines (He II) at 6560.10 and 4859 Å are present in the spectra but are not shown in this table, since the line profiles are contaminated by H$\alpha$ and H$\beta$, respectively, and we make no line-width measurements.

We calculate a redshift of $z = 0.000869644 \pm 0.000042$ by measuring the centroids of Na I D absorption and Ca II absorptions. Our recessional velocity is $20 \pm 12\ \mathrm{km\ s^{-1}}$ greater than the reported heliocentric radial velocity of $241\ \mathrm{km\ s^{-1}}$ (G. de Vaucouleurs et al. 1995) from H I observations of M101 (corresponding to $z = 0.000804$). Figure 2 shows the identification of the Milky Way, M101, and SN 2023ixf Na I D and Ca II K components. There is a strong absorption component at our adopted redshift in both lines. Since we use normalized spectra for the majority of our analysis, we make no correction for interstellar reddening. All spectra and emission lines are reported in air wavelengths.

### 2.2. Photometry

In order to interpret our spectra in terms of the bolometric output of the SN, we obtained the bolometric light curve published in E. A. Zimmerman et al. (2024). From this light curve, shown in Figure 3, we measure the total radiated energy at various epochs. This is discussed more in Section 3.4. To supplement the bolometric light curve, we also retrieved publicly available photometry from Chuck Cynamon (CCHD),[12] Franky Dubois,[13] and Ray Tomlin[14] from The American Association of Variable Star Observers (A. A. Henden et al. 2016). CCHD images were calibrated using the ACP Observatory Control Software. Aperture photometry was applied to SN 2023ixf using AAVSO's VPHOT software. This technique also applied transform coefficients to the observations, so measurements are in Johnson *BVRI*. These data are shown in Figure 3. This plot includes the discovery observation by K. Itagaki (2023) and the last good-quality pre-SN point (C. L. Ransome et al. 2024). Also marked in this figure are the epochs for which we obtained NEID spectra.

[11] NIST Atomic Spectra Database (https://www.nist.gov/pml/atomic-spectra-database).

[12] C. Cynamon: https://aavso.org/users/ccynamon.

[13] F. Dubois: https://www.aavso.org/users/astrosun.

[14] R. Tomlin: https://www.aavso.org/users/tre.

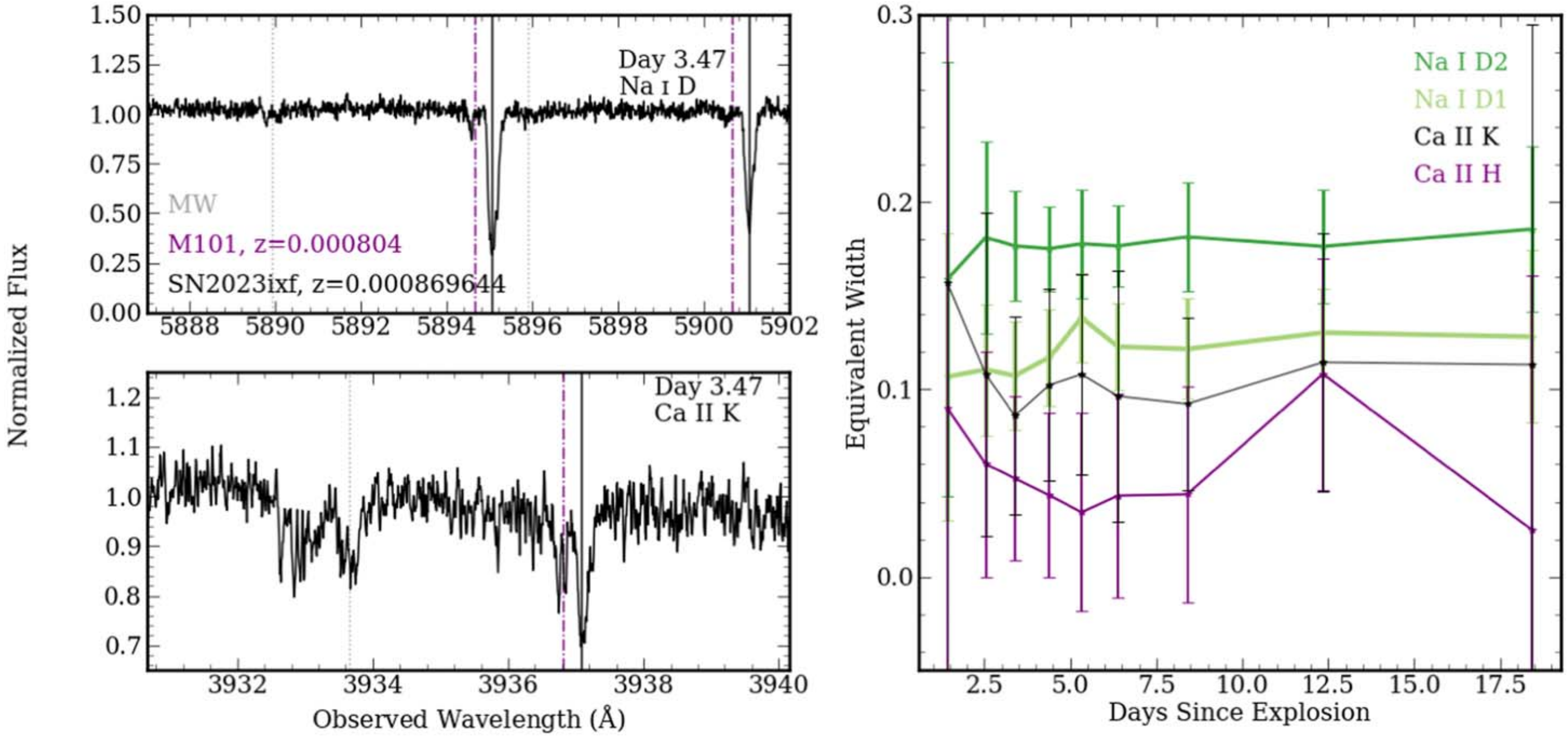

**Figure 2.** Left: Na I D and Ca II K spectra on day 3.47. The wavelengths are as observed by NEID. Milky Way absorption is denoted by the gray dotted line. Absorption corresponding to the redshift of M101 is shown with a purple dotted–dashed line. The solid black line denotes the redshift of SN 2023ixf that we measured. Right: equivalent width measurements of Na I D and Ca II H&K lines observed from the host galaxy.

## 3. Results

### 3.1. Na I D and Ca II H&K

Na I D and Ca II H&K are commonly used to calculate the reddening due to the interstellar medium along the LOS. However, a few SNe have shown temporal evolution of these features, which may indicate that these lines can originate from the CSM. The Type Ia SN 2006X exhibited time variability in its Na I D absorption due to changes in the ionization state of the CSM (F. Patat et al. 2007). SN 2011A, a SN IIn, showed low-velocity Na I D absorption that increased in strength by a factor of $\sim$3 in the months following explosion (T. de Jaeger et al. 2015). These observations were accompanied with narrow H$\alpha$ and a P Cygni profile with a velocity consistent with the velocity of the Na I D, 1100 km s$^{-1}$, showing a shared origin in the CSM.

We measured the equivalent width (EW) at all epochs of Na I D$_1$, Na I D$_2$, and Ca II H&K absorption with the IRAF `splot` function, and these data are shown in Figure 2. There is minimal evolution of Na I D$_1$ and D$_2$, with an overall average EW of 0.176 $\pm$ 0.042 Å that is in agreement with S23. The Ca II H&K lines are also consistent with negligible change within measurement uncertainties. The FWHM of Ca II and Na I D for all epochs of spectra remain constant $\sim$10 km s$^{-1}$. The ionization of the CSM, which excites He II, N III, C III, N IV, and C IV in our spectra, is much too high for any Na I D to form. This, combined with the lack of line profile evolution, suggests the absorption of Na I D and Ca H&K in our spectra is unrelated to the CSM.

**Table 1**
NEID Spectra Observations

| Epoch | MJD | Phase[a] |
| --- | --- | --- |
| 20/05/2023 | 60084.26 | 1.51 |
| 21/05/2023 | 60085.37 | 2.62 |
| 22/05/2023 | 60086.22 | 3.47 |
| 23/05/2023 | 60087.22 | 4.47 |
| 24/05/2023 | 60088.18 | 5.41 |
| 25/05/2023 | 60089.21 | 6.46 |
| 27/05/2023 | 60091.23 | 8.48 |
| 31/05/2023 | 60095.18 | 12.43 |
| 06/06/2023 | 60101.26 | 18.51s |

**Note.**
[a] Phase is with respect to the date of explosion, MJD = 60082.75.

### 3.2. Narrow Emission 1.51 days after Explosion

The lines present in our spectrum are consistent with previously published, early low-resolution ($R \lesssim 3000$; $\gtrsim$100 km s$^{-1}$) spectra (W. V. Jacobson-Galán et al. 2023; K. A. Bostroem et al. 2024). However, our superior kinematic resolution of order 10 km s$^{-1}$ allows for detailed inspection of emission-line profiles. Our 1.51 day spectrum is the earliest high-resolution spectrum ever obtained for SN 2023ixf, and subsequent epochs overlap with S23, starting on day 2.62, which we compare to throughout the paper.

Several profiles of narrow lines with various ionization states are overplotted in Figure 4. The spectra have been continuum-subtracted and peak-normalized, so the line profile shapes can be compared. All transitions plotted are singlets except for the following:

1. He I 5875.62 Å: lines at 5875.61397 Å, 5875.6148 Å, 5875.62510 Å, 5875.6404 Å, and 5875.9663 Å.

**Table 2**
Identification of the Narrow, CSM-formed Emission Lines in Our Spectra

| Wavelength[a] (Å) | Line | 1.51 days FWHM ($km\ s^{-1}$) | 2.62 days FWHM ($km\ s^{-1}$) | 3.47 days FWHM ($km\ s^{-1}$) |
|---|---|---|---|---|
| 6562.81 | H$\alpha$ | 70 (3)[b] | 130 (5) | 250 (17) |
| 4861.30 | H$\beta$ | 80 (4) | 200 (4) | 200 (4) |
| 4340.47 | H$\gamma$ | 50 (4) | … | … |
| 4101.73 | H$\delta$ | 50 (3) | … | … |
| 3970.00 | H$\epsilon$ | 50 (4) | … | … |
| 5875.62[c] | He I | 25 (2) | … | … |
| 6678.15[f] | He I | 50 (3) | … | … |
| 7065.19[f,d] | He I | 40 (2) | … | … |
| 5411.52 | He II | 80 (2) | 300 (3) | 300 (3) |
| 4685.57[e] | He II | 70 (4) | 200 (10) | 400 (16) |
| 4634.14 | N III | 60 (6) | … | … |
| 4640.64 | N III | 60 (7) | 100 (1) | … |
| 4647.42 | C III | 60 (6) | … | … |
| 4650.25 | C III | 60 (2) | … | … |
| 7109.35[f] | N IV | 181 (14) | 330 (9) | … |
| 5801.33[f] | C IV | 250 (17) | 400 (12) | 500 (14) |
| 5811.98[f] | C IV | 200 (8) | 400 (8) | … |
| 7122.98[f] | N IV | 181 (9) | 300 (6) | 700 (14) |

**Notes.** The FWHM of the functional comparisons are listed along with which function was used. Empty entries indicate epochs where a line was not observable. Functional comparisons are shown in Appendix Figures 11, 12, and 13.
[a] Lines are from the NIST Atomic Spectra Database (https://www.nist.gov/pml/atomic-spectra-database).
[b] The $1\sigma$ error associated with the noise in the spectra is shown in parentheses.
[c] He I multiplet: 5875.61397 Å, 5875.6148 Å, 5875.6251 Å, 5875.6404 Å, 5875.9663 Å.
[d] He I triplet: 7065.1771 Å, 7065.2153 Å, 7065.7086 Å.
[e] He II multiplet (only strong lines have been listed): 4685.3769 Å, 4865.4072 Å, 4685.7039 Å, 4685.7044 Å, 4685.8041 Å.
[f] Linear functions were subtracted from these profiles instead of a $\sim$1500 $km\ s^{-1}$ Gaussian.

2. He I 7065.19 Å: strongest lines at 7065.1771 Å, 7065.2153 Å, and 7065.7086 Å.
3. He II 4685.57 Å: strongest lines at 4685.3769 Å, 4865.4072 Å, 4685.7039 Å, 4685.7044 Å, and 4685.8041 Å.

The choice of rest wavelength for 5875.62 and 7065.19 Å is the observed wavelength of the strongest lines reported by NIST. The rest wavelength for 4685.57 Å is an average of the strongest lines listed here. With the exception of H$\alpha$, He I 5875.62 Å, and the He II 4686 Å blend, all lines in this first epoch are blueshifted with peaks around $-40\ km\ s^{-1}$. These values are smaller than the $-50$ to $-150\ km\ s^{-1}$ range observed in the day 2.6 spectra presented by S23.

In this first epoch, line widths vary significantly with ionization potential. The following line velocity measurements are the speeds of the CSM in the region where a given ion is present and where emission lines form. We find that He I 5875.62 Å is the narrowest line, with a FWHM $= 25 \pm 2\ km\ s^{-1}$. The error is $1\sigma$, where $\sigma$ is the standard deviation in the spectrum continuum. We adopt this value as the speed of the CSM region where He I lines form on day 1.51. We discuss the error measurement in Section 3.4. H I, He II, C III, and N III have FWHM values between 60 and $80\ km\ s^{-1}$, whereas C IV and N IV have FWHM values between 180 and $250\ km\ s^{-1}$. The widths of all lines were estimated using the blue half of the line profile ($v < 0\ km\ s^{-1}$), since we assume the red side is blocked by high optical depth (J. H. Groh 2014; I. Shivvers et al. 2015).

For N III/IV and C III/IV, the line profiles of each transition exhibit consistent overlap. The He II lines have considerably different profiles, reflecting the multiplet around the

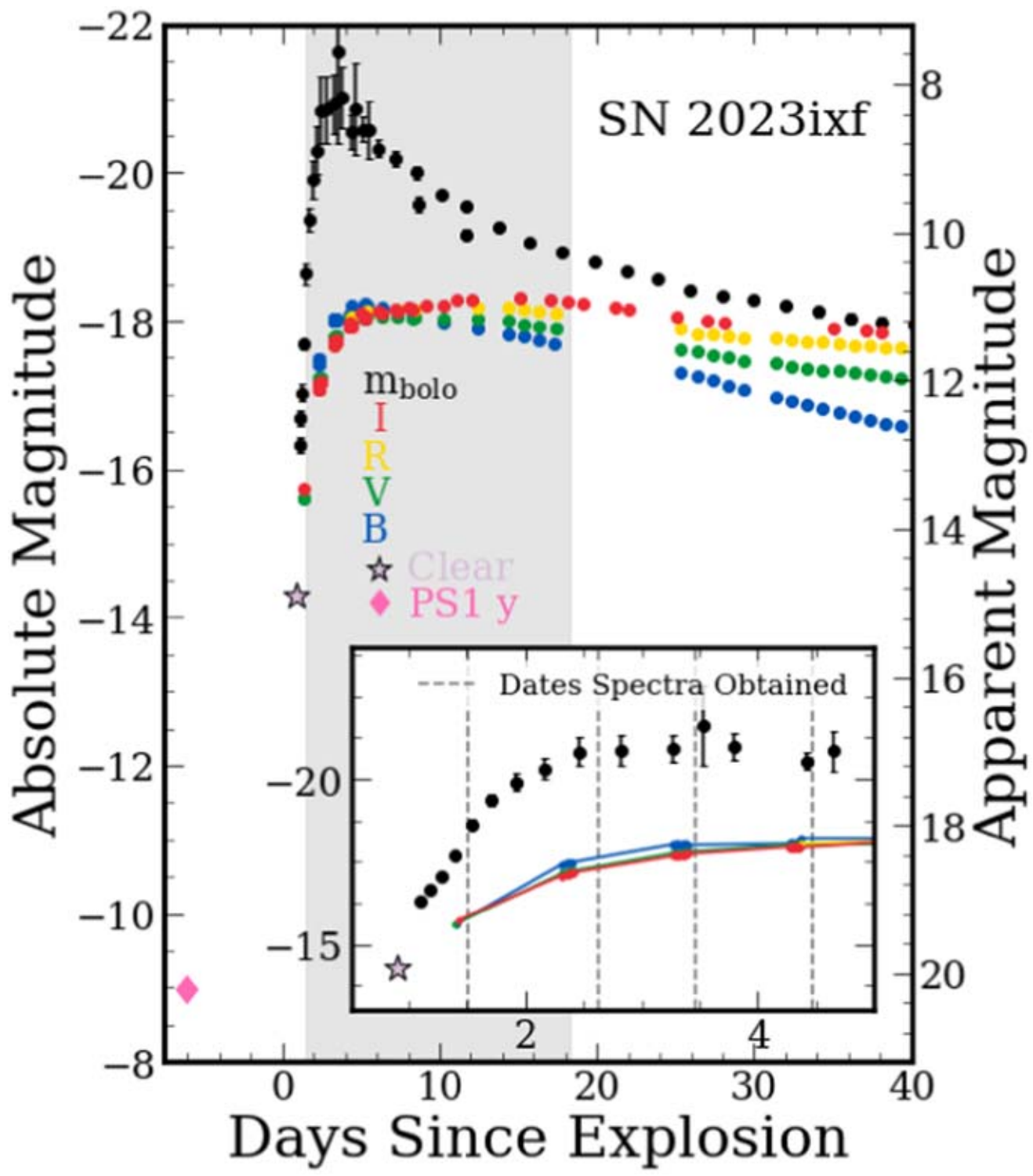

**Figure 3.** AAVSO *BVRI* photometry of SN 2023ixf. The bolometric light curve derived from Swift, the Hubble Space Telescope, and ground-based telescopes is overplotted (E. A. Zimmerman et al. 2024). The purple star is the discovery detection by K. Itagaki. The pink diamond is the last good-quality pre-SN point presented in C. L. Ransome et al. (2024). The shaded region indicates the window in which NEID spectra were taken. The first four epochs of spectra are shown as vertical dashed lines in the inset.

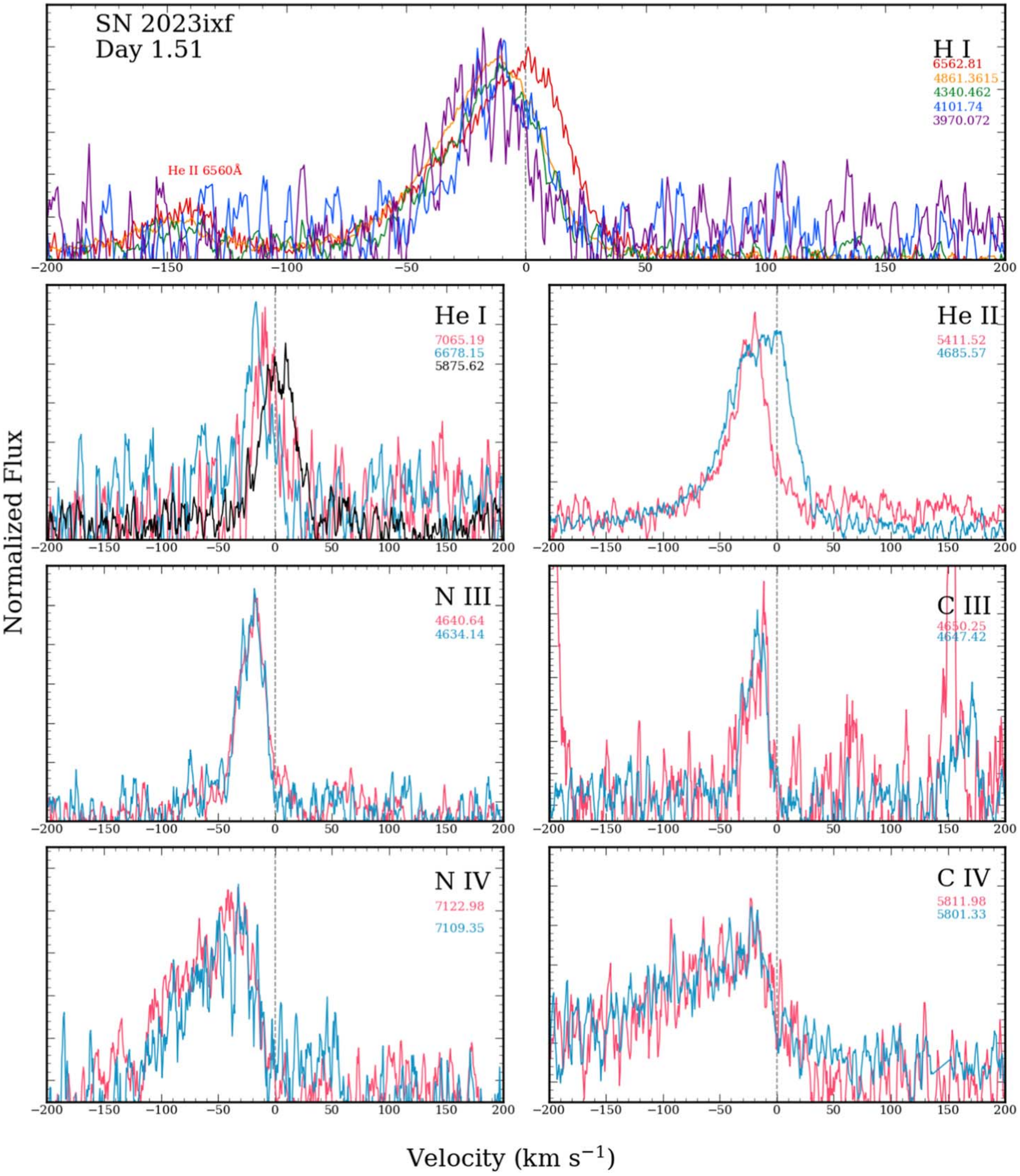

**Figure 4.** Doppler velocity plots of various ions. Velocities are with respect to the rest wavelength listed on each panel. All spectra have been continuum-subtracted. All lines have been convolved with a 0.4 Å filter, except for Hα (no convolution).

4685.57 Å line. The Hα is unlike the rest of the Balmer series, as it has a relatively symmetric profile. We do not observe a narrow P Cygni profile in any emission line at this epoch, consistent with echelle spectra in subsequent days presented by S23. All three line profiles of He I are dissimilar from one another.

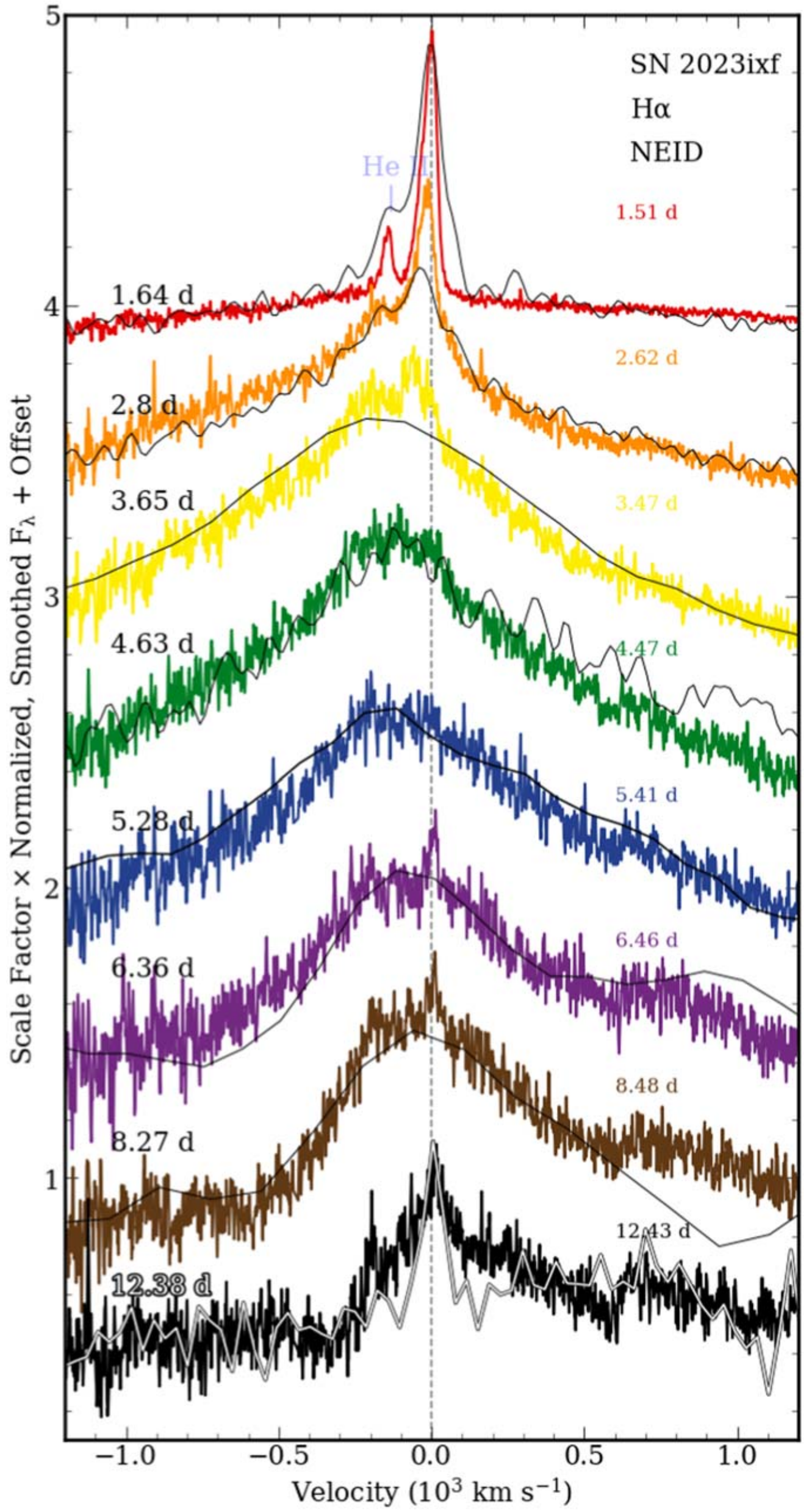

**Figure 5.** Evolution of Hα over the first 12 days. Spectra are not corrected for extinction. Colorful lines are the WIYN/NEID echelle spectra, and the overlaid black lines are low-resolution spectra at similar timescales. Continuum-subtracted spectra are normalized by their peak emission and separated by an offset. Spectra have been convolved to a 0.5 Å filter.

### 3.3. Time Evolution of Narrow Lines

The narrowest components of emission lines are visible in our spectra for ∼4 days before disappearing. Broader features persist in the spectra (N. Smith et al. 2023) but span velocities that overlap with the edge of the orders and cannot be analyzed due to poor signal-to-noise ratio. In this section, we discuss the time evolution of these features relative to the day 1.51 spectrum.

The narrow He I feature vanishes after the first epoch. S23 do not observe He I emission in the LBT/PEPSI spectra until it reappears (very weakly, and only in the intermediate-width component) on day 8.48. In the NEID spectra, noise dominates over this weak line, so we do not observe He I after day 1.51. Otherwise, the spectral evolution presented here is very similar to the evolution in S23. J. Zhang et al. (2023) observe He I 6678.15 Å on day 1.2, and it vanishes by day 1.8.

Figure 5 illustrates the rapid evolution of Hα between days 1.51 and 12.43. Overplotted in black are publicly available low-resolution spectra (see also J. Zhang et al. 2023; K. A. Bostroem et al. 2024).[15] The EW of the narrow-line component drops by a factor of 10 from day 1.51 to 2.62. The normalized profile broadens from a 70 km s$^{-1}$ feature to 350 km s$^{-1}$ over the first 4 days. During this broadening, the centroid of Hα shifts from 0 km s$^{-1}$ to blueshifted velocities, peaking at -200 km s$^{-1}$ after 4 days.

There is a marginal narrow emission component of Hα observed on and after day 6.46. This component persists in its mean velocity (15 km s$^{-1}$) and its width (∼30 km s$^{-1}$) until at least day 17.56, which was observed by S23 as an emission line centered on 0 km s$^{-1}$ with a FWHM of 45 km s$^{-1}$. The values we measure are consistent with those reported in S23 within uncertainty. The emission could be associated with an adjacent H II region. It could also arise from a distant, photoionized RSG wind.

Up to four epochs of several strong narrow emission lines are shown in Figure 6. Balmer transitions start with narrow profiles that broaden and weaken with time. The peak of the Hα emission line is at 0 km s$^{-1}$ on day 1.51, with tails that extend from −100 to +15 km s$^{-1}$. The center of the line moves blueward as the line broadens, and the narrow component is gone after day 4.47. Hβ, Hγ, Hδ, and Hϵ initially peak at −20 km s$^{-1}$, and then broaden like Hα but fade more quickly. He II and N III have similar evolution to the Balmer series, with a width of ∼70 km s$^{-1}$ on day 1.51 that broaden and fade. N IV and C IV begin with very asymmetric, blueshifted profiles with blue wings extending to 200 km s$^{-1}$. The blueshifted sides of the lines also broaden with time before they completely fade away. C III and N III rapidly fade over the first 2 days, whereas N IV and C IV persist until day 4. C III and N III have symmetric profiles with centroids shifted by −20 km s$^{-1}$, and they disappear after days 1.51 and 2.62, respectively. The full time evolution of these lines is presented in the Appendix. The radiation emitted from continual shock–CSM interaction further ionizes the gas into N IV, C IV, and He II. Once the shock overcomes all CSM, then the gas cools and recombines, which is suggested by the reappearance of He I (S23).

### 3.4. Radiative Acceleration

The interaction between the SN shock and CSM produces UV and optical photons that ionize and can also accelerate the CSM. The time evolution of the pre-shock CSM speed, and thus the width of the line profiles we observe, is strongly affected by radiative acceleration. The maximum speed that Thomson scattering can accelerate spherically symmetric CSM (external from the shock) is

$$\Delta v_{\rm CSM} = \frac{80 E_{49}}{r_{15}^2}\ {\rm km\ s^{-1}}, \tag{1}$$

where $E_{49}$ is the total radiated energy in units of $10^{49}$ erg, and $r_{15}$ is the radius of CSM in units of $10^{15}$ cm (N. N. Chugai et al. 2002).

We integrated the bolometric light curve from the time of explosion to the first epoch of spectroscopy and estimate that the SN emitted ≈6.4 × $10^{46}$ erg. We adopt a shock velocity of 9500 ± 500 km s$^{-1}$, estimated from a publicly available spectrum from WISeREP obtained with the Dark Energy Spectroscopic Instrument (DESI Collaboration et al. 2016) mounted on the Mayall 4 m Telescope at KNPO on MJD = 60116.146, 33.4 days following explosion. This measurement is the bluest edge of the broad P Cygni component in H$\alpha$, where the blue wing fades into the continuum at $-9500$ km s$^{-1}$ and represents a lower limit to the shock velocity. It is likely that the shock decelerated as it propagated through the CSM, but we do not observe a broad P Cygni component at early times (<5 days) from which we can measure the shock velocity. The associated radius of the shock at 4.47 days is $3.7 \times 10^{14}$ ($v_{\rm shock}/9500$ km s$^{-1}$) cm, where $v_{\rm shock}$ is the velocity of the shock.

[15] WISeREP: https://www.wiserep.org/object/23278.

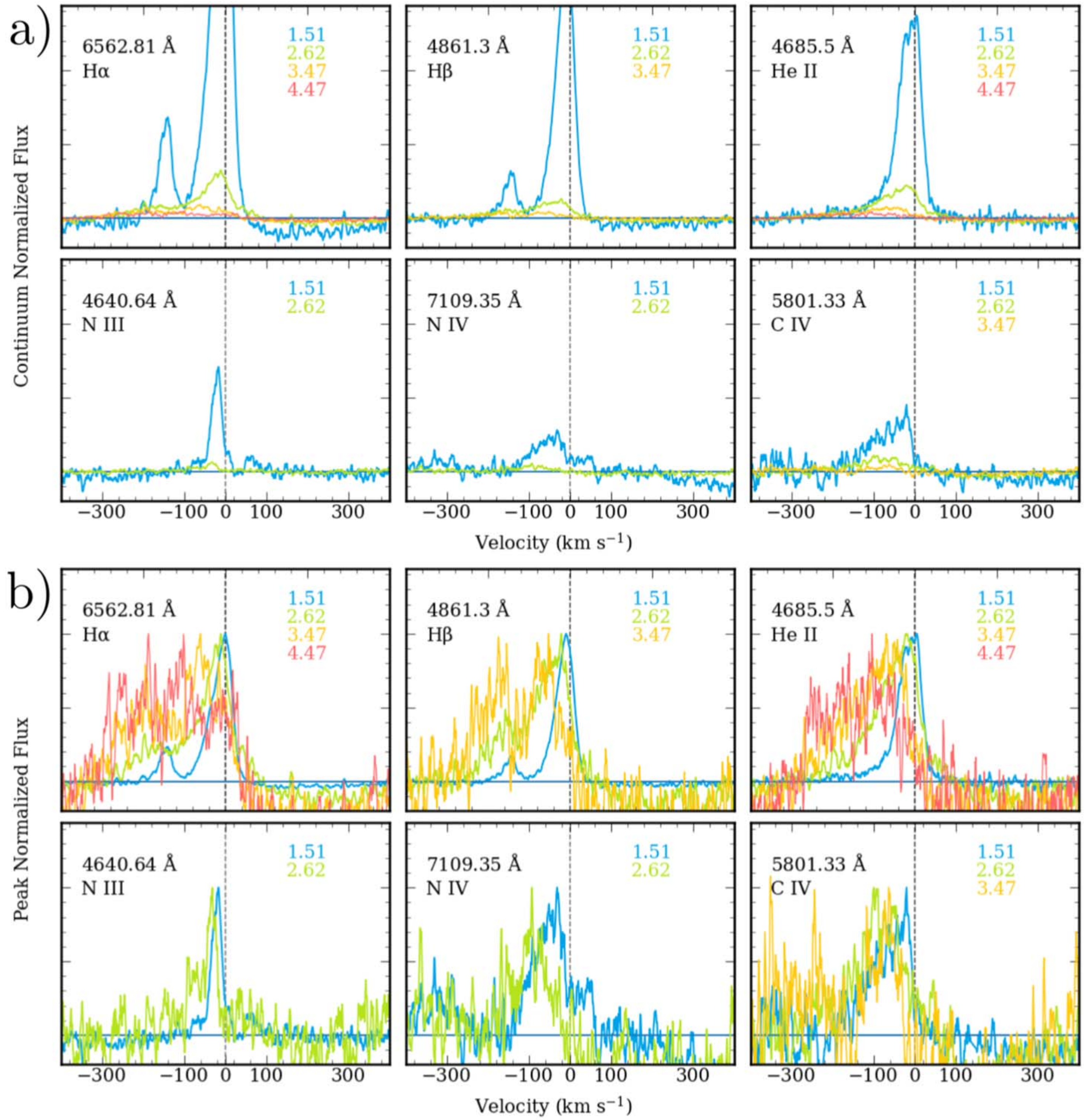

**Figure 6.** Continuum-subtracted spectra of several emission lines over the first four epochs. Spectra have not been corrected for extinction, and they have been convolved with a 0.83 Å filter. Only epochs with discernible emission are plotted. (a) Spectra have been continuum-subtracted. (b) Spectra have been continuum-subtracted and peak-normalized.

From Equation (1), we calculate that the CSM immediately ahead of the shock ($r = v_{\rm shock} \times t$) could have been accelerated to a maximum speed of 45 (300, 500) km s$^{-1}$ by day 1.51 (and by days 2.62 and 3.62). These values mildly disagree with the broadening of narrow emission lines we observe. Quantifying radiative acceleration of the CSM is complicated by the fact that optical depth and the amount of radiative acceleration are dependent on proximity to the forward shock. At high optical depths, the CSM could have absorbed more momentum because of multiple scatterings, thereby increasing the radiative acceleration. On the other hand, observations may be more sensitive to regions with lower optical depth and would be subject to less radiative acceleration.

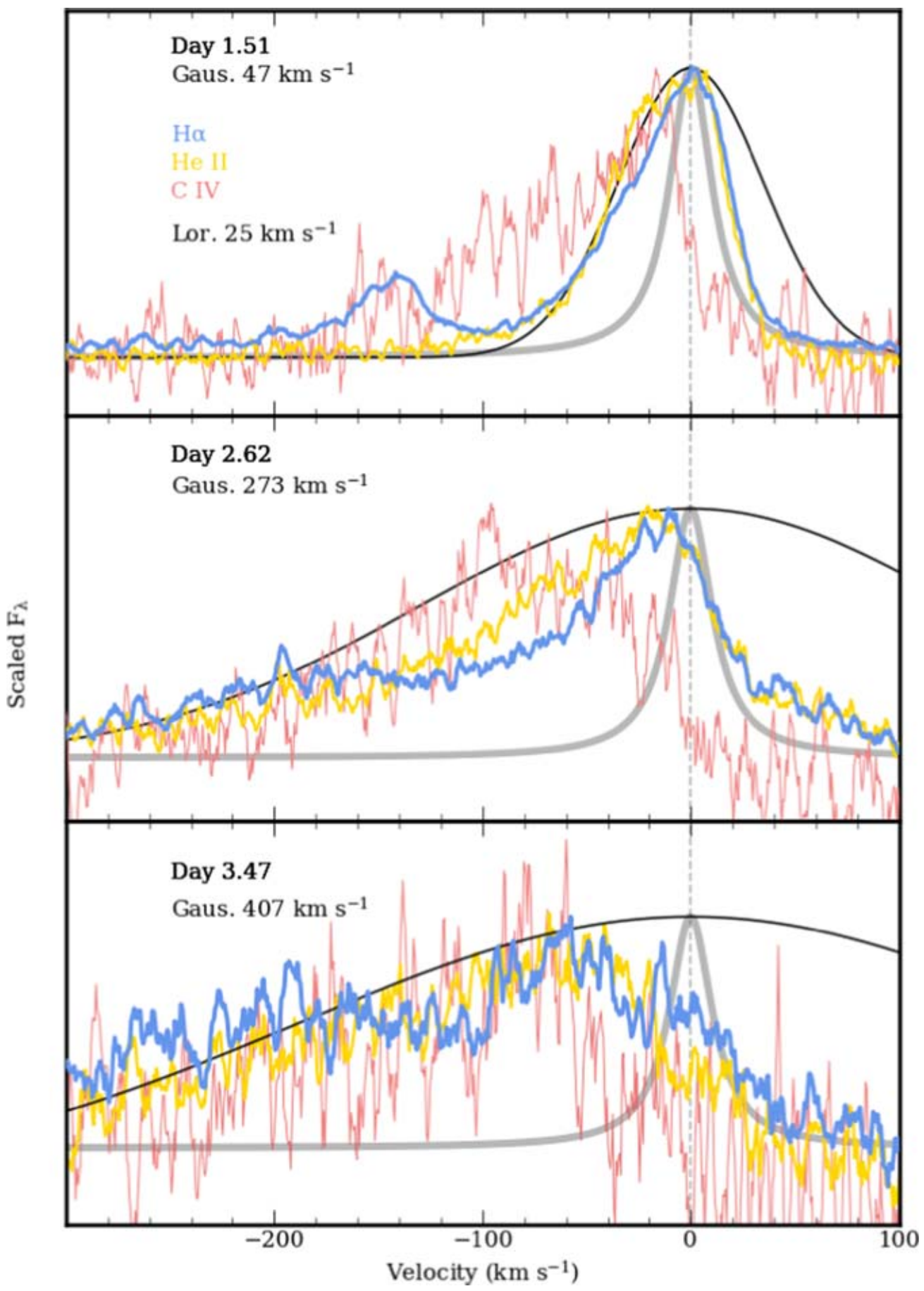

**Figure 7.** Comparison of Hα, He II 4685.5 μm, and C IV 5801.33 μm over the first 3 days. Spectra are convolved with a 0.6 Å filter. Overplotted in each panel are a Lorentzian with a FWHM = 25 km s$^{-1}$ and a Gaussian with a FWHM that reflects the maximum speed Thomson scattering could accelerate the CSM to at each epoch.

The outer edge of the dense CSM/CSM farthest from the shock ($3.7 \times 10^{14}$ ($v_{\rm shock}$/9500 km s$^{-1}$)) could have been accelerated by a maximum speed of 5 km s$^{-1}$ by day 1.51. Hence, there is an uncertainty of 0–5 km s$^{-1}$ in the amount of radiative acceleration at the speed of 25 km s$^{-1}$. There is also a measurement error of 2 km s$^{-1}$ in the continuum of the spectra. Since this is the narrowest line in our spectra, we interpret this to indicate that at least some material within $r_{\rm CSM}$ had a pre-SN speed of $25^{+0}_{-5} \pm 2$ km s$^{-1}$. In reality, the CSM likely had a complex density and velocity structure. For instance, the other He I lines, at 6678.15 and 7065.19 Å, have speeds of 50 and 40 km s$^{-1}$, respectively, faster than the 25 km s$^{-1}$ speed we measure from the 5875.62 Å line.

The progressive radiative acceleration of the CSM can be observed in the time evolution of line profiles with varying ionization energies. Figure 7 shows the time evolution of Hα, He II 4685.5 μm, and C IV 5801.33 μm, which is similar to Figure 9 in S23 but with the addition of our spectrum on day 1.51. Each epoch is plotted with a Lorentzian with a FWHM of 25 km s$^{-1}$, which corresponds to the width measured from He I. Each epoch also shows a Gaussian with a FWHM that corresponds to the maximum radiative acceleration at the shock's location ($r = v_{\rm shock} \times t$).

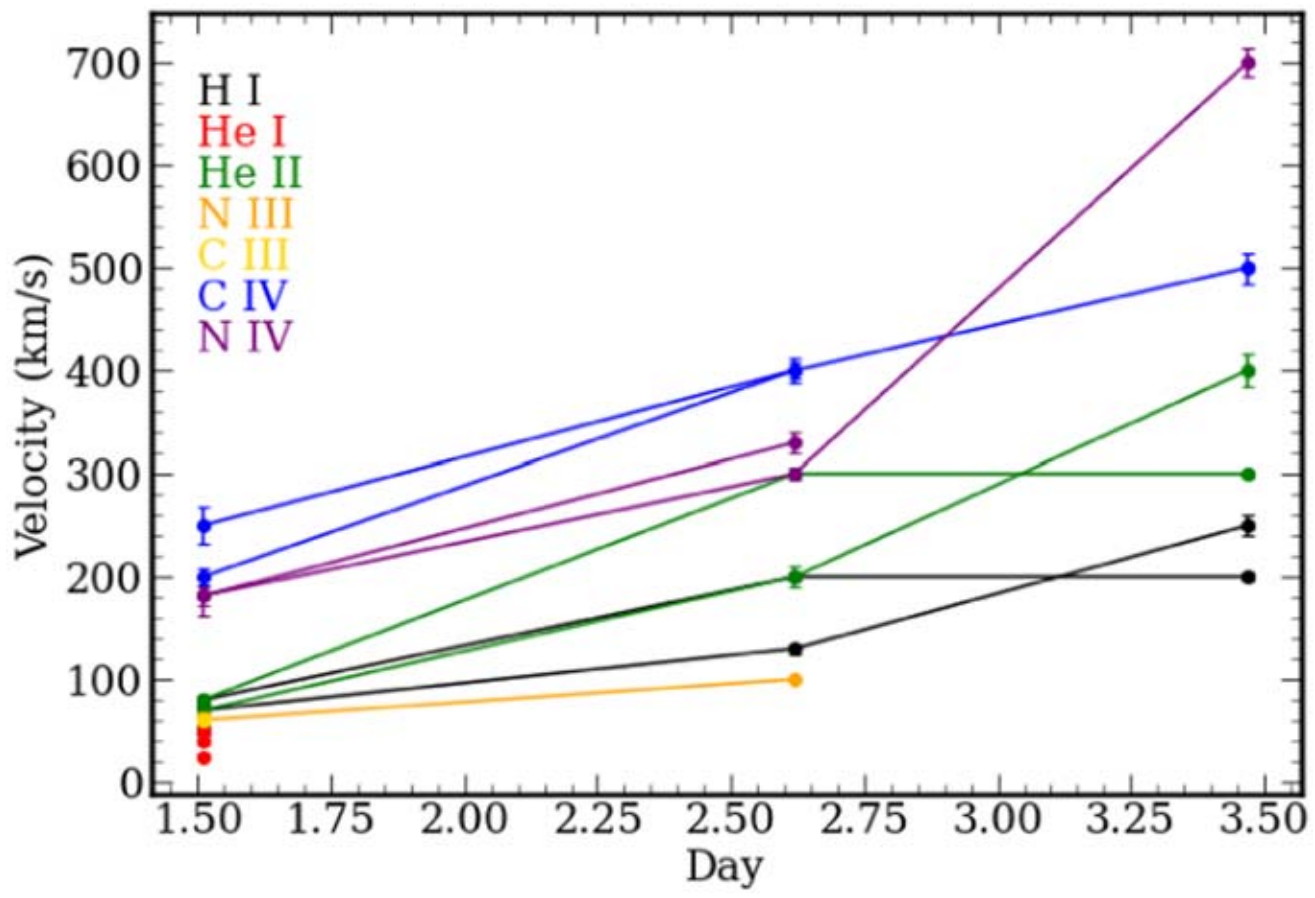

**Figure 8.** Velocities presented in Table 2 plotted against time. H I is shown in black. Black, red, orange, and yellow lines denote lower ionization energies ($E \lessapprox 50$ eV). Green, blue, and purple lines denote higher ionization energies ($50 \lessapprox E \lessapprox 80$ eV).

There is disagreement between the maximum radiative acceleration and the speed of C IV on day 1.51. The first epoch of C IV is faster than the calculated radiative acceleration, which may be a result of the single-scattering assumption made for the equation used here. On the second and third days, there is better agreement between the maximum radiative acceleration and the C IV emission ($\sim$350 km s$^{-1}$), which traces gas closer to the shock. As the bolometric output of the SN increases, CSM further out from the shock is accelerated. For the maximum radiative acceleration measured here ($\Delta v < 500$ km s$^{-1}$), it is clear that the intermediate-width profiles (1000–3000 km s$^{-1}$) originate from shocked CSM or electron scattering, and the narrowest emission ($<$500 km s$^{-1}$) is indicative of pre-shock CSM. We interpret the narrow-component time evolution over the first 4 days to be mostly due to radiative acceleration driven from the SN radiation. The disappearance of narrow-line profiles around 4 days correlates well with the He II flux maximum and the UV peak timing (E. A. Zimmerman et al. 2024), as well as the change in position angle of the polarization (S. S. Vasylyev et al. 2023; A. Singh et al. 2024) on day 4.5.

In Figure 8, we plot the FWHM velocities presented in Table 2 against time since explosion. The data for each ion end when a line is no longer present, and lines of the same color represent different transitions for a single ionization state. Narrow emission in states like He I, N III, and C III vanish after the first epoch, and the increasing ionization in the CSM results in emitting He II, N IV, and C IV until the shock passes through all CSM, consistent with observations by S23. Since ions with higher ionization states last longer and trace the hottest gas, they serve as an excellent probe of the acceleration of the CSM ahead of the shock.

While the pre-shock CSM was asymmetric and probably had a range of outflow speeds, unlike a steady spherical wind, we find that the expected amount of radiative acceleration of pre-shock CSM provides a reasonable explanation for most of the observed time evolution in narrow-line profile width over the first few days after explosion.

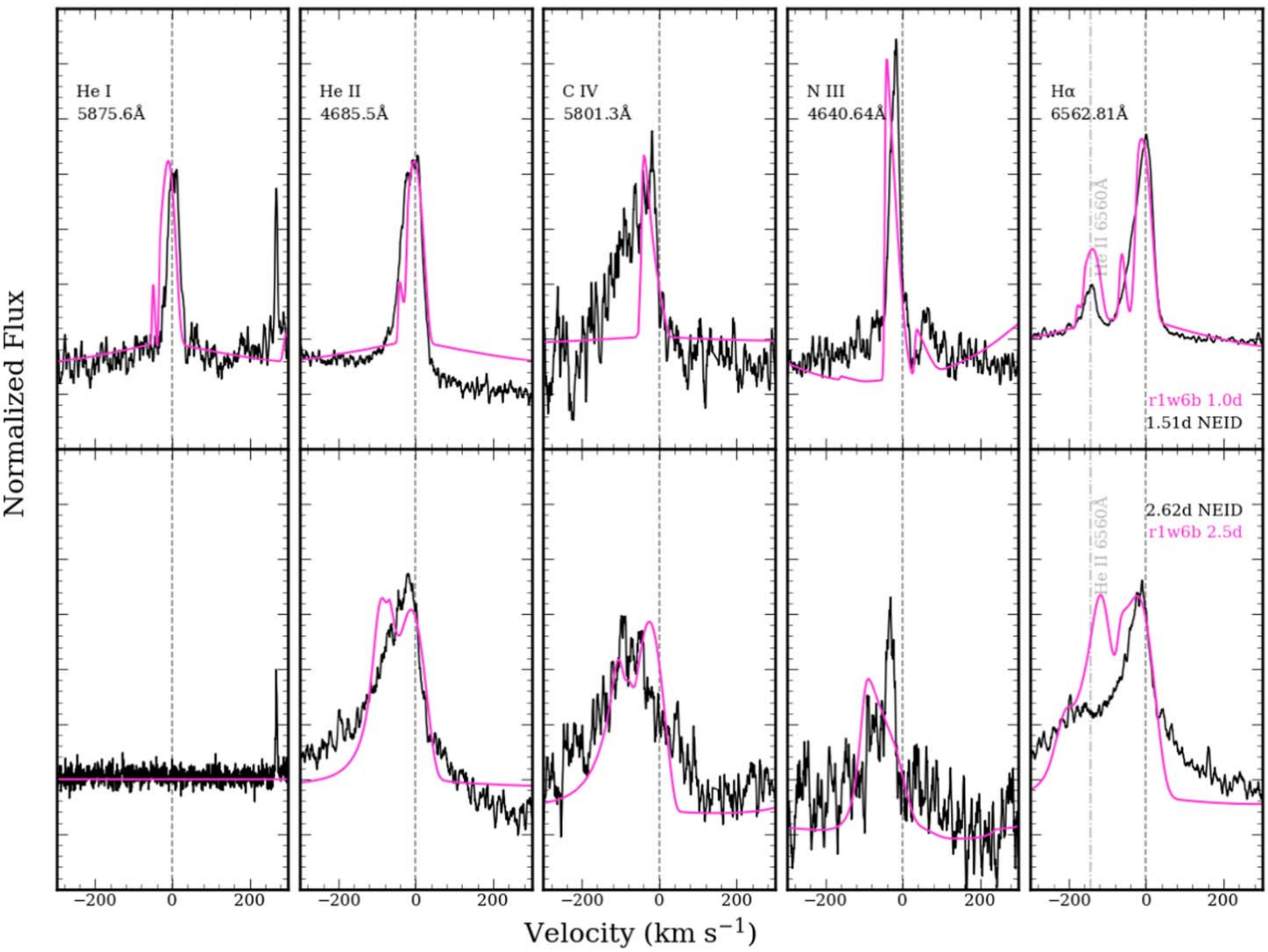

**Figure 9.** The first two epochs of the SN 2023ixf spectra and a prediction by the r1w6b model for Hα, He I, He II, C IV, and N III. All spectra have been continuum-subtracted and peak-normalized. For the second epoch of He I, the model and spectrum have been plotted without peak scaling. The NEID spectra have been convolved with a 0.83 Å filter.

## 4. Discussion

### 4.1. CMFGEN Models

W. V. Jacobson-Galán et al. (2023) use 1D HERACLES/CMFGEN models (D. J. Hillier & L. Dessart 2012; L. Dessart et al. 2015, 2017) to constrain the mass-loss parameters of SN 2023ixf, and find that evolution of the r1w6b model, which is for a $15 M_{\odot}$ RSG with a spherical CSM distribution with $v_{\infty} = 50\,\mathrm{km\,s^{-1}}$, closely matches the intermediate-resolution spectra. In this section, we compare this model with the NEID spectra to constrain the CSM and validate the radiative acceleration interpretation.

We compare the first two epochs of our NEID spectra with a high-resolution ($\sim$0.1 Å, $5\,\mathrm{km\,s^{-1}}$ at Hα) r1w6b model in Figure 9, the model having a mass-loss rate of $0.01\,M_{\odot}\,\mathrm{yr^{-1}}$ (L. Dessart et al. 2017; L. Dessart 2025). The spectra are continuum-subtracted and peak-normalized. The continuum of each model spectrum is fit with a cubic polynomial around $0\,\mathrm{km\,s^{-1}}$ for each line and is subtracted. The emission is then peak-normalized.

The widths and centroid velocities of each narrow emission line of interest are generally in agreement with this model. The Hα profile in the model is centered at $0\,\mathrm{km\,s^{-1}}$ and has line wings that extend from $-100\,\mathrm{km\,s^{-1}}$ to $+20\,\mathrm{km\,s^{-1}}$. There is a P Cygni profile near $-50\,\mathrm{km\,s^{-1}}$, which corresponds to the initial condition $v_{\infty} = 50\,\mathrm{km\,s^{-1}}$. The Hα profile in the day 1.51 spectrum is a similar width but lacks a narrow P Cygni absorption. For ions closer to the shock, like He II, N III, and C IV, the model predicts blueshifted lines with blue wings extending from $-50\,\mathrm{km\,s^{-1}}$ to $-150\,\mathrm{km\,s^{-1}}$ as the CSM is accelerated. The He II and N III spectra match the model well in width, but the fine structure of the lines differs. The observed C IV exhibits broader line profiles than those predicted by the model, but the emission profiles are both blueshifted. Due to the single ionization state in H I, Hα traces a much larger volume of gas and is less influenced by the optical depth effects that suppress emission from the far side of the CSM. The same is true for He I: The strongest line, 5875.62 Å, traces a larger radius, whereas the other lines (N III and C IV) trace the interior of the CSM.

The model's He I 5875.62 Å profile is well matched by a $50\,\mathrm{km\,s^{-1}}$ Gaussian, so the process of comparing a $25\,\mathrm{km\,s^{-1}}$ Gaussian to our 1.51 days NEID spectrum is valid for inferring the pre-SN CSM speed. He I then fades rapidly in the model spectra, exactly as we observe. The line broadening in Hα, He II, C IV, and N IV is reasonably similar, as well, although the relative strengths between the Pickering line and Hα are different.

An important difference to note is that the EW of narrow H$\alpha$ in the r1w6b model does not change from 1 to 2.5 days, where the EW of the narrow H$\alpha$ in the NEID spectra changes by a factor of 10 between days 1.51 and 2.62. This could be due to the shock engulfing a large fraction of CSM between days 1.51 and 2.62, as suggested by S23. This evolution would therefore not occur in the model in this case, since the CSM is a steady wind. Additional models with fluctuations in the CSM density profile would further constrain whether the drop in flux is due to CSM being swept up or due to another mechanism.

A P Cygni profile is predicted for He I, He II, and H$\alpha$ in the model, but this feature is absent in the NEID spectra. The above inference and the disagreements between line profile shapes corroborates the notion that the spectral parameters may be dependent on the geometry of the CSM as well as additional deviations from the smooth wind and the $\sim r^{-2}$ density structure of the r1w6b model. Line profiles also change rapidly during this phase.

S. S. Vasylyev et al. (2023), A. Singh et al. (2024), and M. Shrestha et al. (2025) present imaging polarimetry and spectropolarimetry over the first 2 weeks after SN 2023ixf. The polarization changes rapidly during the first 4 days following explosion. The time evolution in polarization is consistent with the disappearance of all narrow emission after day 3.39, which suggests an asymmetric CSM structure that is overrun by the shock's outflow around 3–4 days after explosion. A candidate geometry is an inclined, equatorial torus of CSM, a system that may be common among massive stars and proposed by S23 for SN 2023ixf. A P Cygni profile would not be observed in this case as there would not be sufficiently dense, absorbing CSM along the LOS.

Differences between the observed spectral evolution and the models may also be due to varying CSM velocity and density profiles. Further modeling of CSM with clumpy density profiles and nonuniform velocities is necessary in order to determine the efficiency of radiative acceleration in these systems and constrain the parameter space where P Cygni profiles will not form.

### 4.2. Parallels to SN 1998S

A high-resolution spectrum of the Type IIn SN 1998S obtained 1.86 days after explosion is presented in I. Shivvers et al. (2015) and merits comparison with our NEID observations of SN 2023ixf. SN 1998S was enshrouded by material that reached up to $10^{17}$–$10^{18}$ cm away from the SN (D. C. Leonard et al. 2000; A. Fassia et al. 2001; L. Dessart et al. 2016). The CSM was constrained to have an inner component ($<10^{15}$ cm) and an outer component. The inner component, the focus of our comparison, was determined to be aspherical but axisymmetric via spectropolarimetry (D. C. Leonard et al. 2000; L. Wang et al. 2001; L. Dessart 2025).

The best-fit CMFGEN model parameters to the high-resolution spectrum of SN 1998S suggest the progenitor underwent enhanced mass loss at a rate of $6 \times 10^{-3}\,M_\odot\,\mathrm{yr}^{-1}$ and with a wind speed of $40\,\mathrm{km\,s^{-1}}$. I. Shivvers et al. (2015) use H$\beta$, H$\gamma$, and He II to constrain these parameters in their day 1.86 spectrum, since these lines trace the inner parts of the CSM. This episode occurred 15 yr prior to explosion, and this dense mass only extends to $\sim 10^{15}$ cm. D. C. Leonard et al. (2000) proposed that the progenitor underwent two strong mass-loss episodes. The first was ejected 20–170 yr prior to explosion and located in a large disk, and the second was ejected within 9 yr of the SN. The wind density profile seems to be similar to the density profile proposed by E. A. Zimmerman et al. (2024) for SN 2023ixf in that a stronger wind "turns on" close to the time of stellar death and is preceded by a weaker wind evident from the late-phase UV observations (A. Singh et al. 2024; K. A. Bostroem et al. 2024).

We compare emission lines for both SNe 1998S and 2023ixf in Figure 10. A value of $z = 0.002915$ is used to correct for the redshift of SN 1998S, determined from the center of the Ca II H&K absorption. Spectra of SNe 2023ixf and 1998S at this early time after explosion exhibit many similarities.

In the two left panels of Figure 10, the narrowest components of H$\alpha$ and He I are nearly identical. The widths of the lines are the same, perhaps suggesting a similar pre-shock wind speed. The only major difference is that SN 2023ixf lacks P Cygni absorption in H$\alpha$.

Neither lines of N IV nor C IV were seen in SN 1998S, unlike SN 2023ixf. The He II and N III profiles differ greatly between SN 1998S and SN 2023ixf, where the He II and N III emission lines are much faster. The differences in kinematics may be attributable to different explosion energies or CSM density profiles and extent. The different velocities and the difference in H$\alpha$ P Cygni may also be due to LOS effects associated with asymmetry in the CSM geometry.

### 4.3. The Progenitor and the Mass-loss Episode

Our observations support the view that broader emission of higher-ionization lines traces gas closer to the shock, where it is hotter. There is likely a velocity gradient in the CSM that is dominated by radially dependent radiative acceleration, as observed by the variable widths of lines in the spectra in comparison with the r1w6b model. In this picture, the narrowest lines are emitted farthest from the shock front, since this acceleration obeys a nearly inverse-square law. This phenomenon is also observed in the CMFGEN r1w6 model; see Figure A.1 in L. Dessart et al. (2017).

Simultaneously, multiple lines of evidence support the interpretation that the CSM was asymmetric. Mild asymmetry associated with the CSM having an internal density contrast of a factor of a few could provide polarization signatures detected by S. S. Vasylyev et al. (2023), as shown in L. Dessart 2025. Additionally, the drop in EW in narrow H$\alpha$ after day 1.51 suggests the shock engulfed a large fraction of CSM.

The CSM geometry could be an equatorial overdensity that decreases away from the plane of the CSM, as proposed by S23, A. Singh et al. (2024), and M. Shrestha et al. (2025). Enhanced mass loss in the years prior to the SN created a torus of CSM, likely from binary interaction with a companion (N. Smith et al. 2003; N. Smith & W. D. Arnett 2014). Massive stars are frequently in binary systems (H. Sana et al. 2012), and recent studies have also invoked prior binary mass stripping to explain the light-curve shape and envelope mass of SN 2023ixf (B. Hsu et al. 2024). In this scenario, the inclined CSM was not visible along the LOS, which would not produce a narrow P Cygni profile. Due to the drop in density away from the plane of CSM, material in lower-density regions may be accelerated more. At later times (>8 days), post-shock CSM absorbs blueshifted light along the LOS, causing the P Cygni profile observed in H$\alpha$ at $-1300\,\mathrm{km\,s^{-1}}$. However, radiative acceleration is independent of density, so another mechanism must be invoked to explain the off-equator acceleration of material.

It is unclear whether this mass loss was due to a single mass-loss event or a longer-lived "superwind" phase of continuous mass loss. The intensified mass loss began $\gtrsim 4\,(v_{\mathrm{shock}}/9500\,\mathrm{km\,s^{-1}})$ years prior to eruption. The SN shock

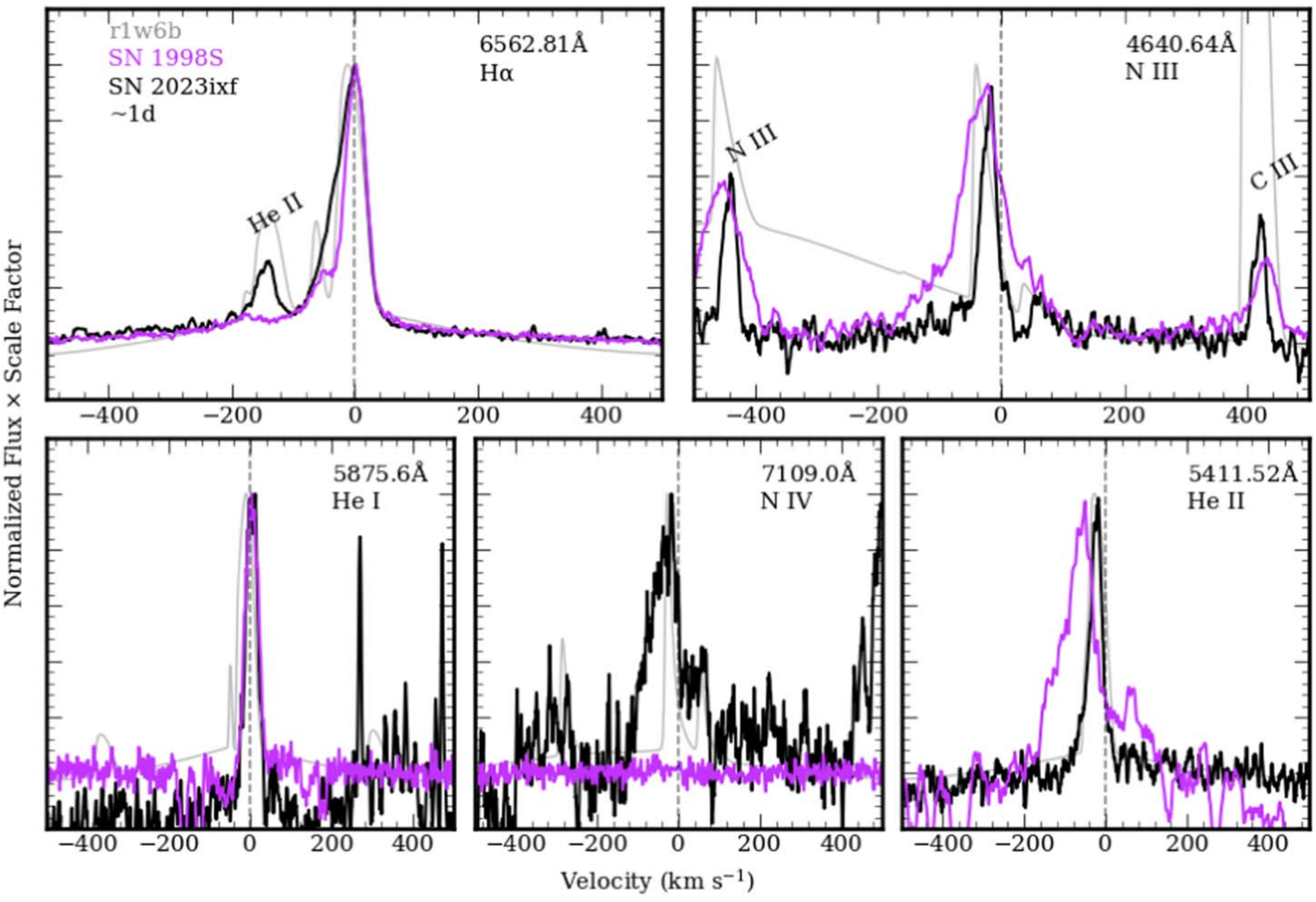

**Figure 10.** Comparison of SN 2023ixf to SN 1998S. Each panel compares the first day of each object's evolution. Spectra have been continuum-subtracted and peak-normalized in both objects. The SN 2023ixf spectra have been convolved with a 0.83 Å filter.

**Table 3**
Reported CSM Parameters Derived for SN 2023ixf

| References | Epoch (days) | $v_{CSM}$ (km s$^{-1}$) | log($r_{CSM}$) (cm) | log($\dot{M}$) ($M_\odot$ yr$^{-1}$) | log($\rho_{CSM}$) (g cm$^{-3}$) | Method |
|---|---|---|---|---|---|---|
| W. V. Jacobson-Galán et al. (2023) | 1.1–13.33 | 50 | 14.00 | −2.00 | −12.00 | Spectral models |
| E. A. Zimmerman et al. (2024) | 2.75 | 30 | 14.28 | −1.96 | −12.30 | Narrow-line centroids |
| J. E. Jencson et al. (2023) | <−1000 | 10 | >14.60 | −3.82 | <−14.32 | Prog. SED model |
| A. J. Nayana et al. (2024) | 4–165 | 25 | 14.60–16.15 | −4.00 | −14.89–17.99 | X-ray and radio modeling |
| J. M. M. Neustadt et al. (2024) | <−500 | 10 | 14.69 | −5.00 | −15.70 | Prog. SED model |
| A. Singh et al. (2024) | 1–120 | 10 | 14.69 | −2 | −12.67 | Light-curve model |
| K. A. Bostroem et al. (2024) | 6.5 | 150 | 14.70 | −2.30 | −13.25 | Spectral models |
| D. Hiramatsu et al. (2023) | 0.08–30 | 115 | 14.70 | −0.30 | −12.18 | Light-curve models |
| P. Chandra et al. (2024) | 13 | 115 | 14.75 | −3.25 | −15.11 | 3–10 keV L, X-ray |
| G. Li et al. (2024) | 0–2.5 | 100 | 14.78 | −1.30 | −13.13 | Light-curve model |
| J. Zhang et al. (2023) | 1.79 | 55 | 14.85 | −3.22 | −15.01 | Hα flux |
| R. S. Teja et al. (2023) | 1–20 | 10 | 14.85 | −3.00 | −14.00 | Light-curve models |
| L. Martinez et al. (2024) | 1.9–18.9 | 115 | 14.90 | −2.52 | −14.66 | Light-curve model |
| B. W. Grefenstette et al. (2023) | 4.4 | 50 | <15.00 | −3.52 | >−15.18 | H I absorption, X-ray |
| M. D. Soraisam et al. (2023) | −6000–0 | 50 | 15.00[a] | −3.52 | −15.52 | Progenitor Period-Luminosity relation |
| Z. Niu et al. (2023) | <−1300 | 115 | 15.18 | −3.70 | −16.417 | Prog. SED model |
| S. Panjkov et al. (2024) | 0.67–3.36 | 50 | 15.60 | ⩽−3.30 | ⩽−16.78 | X-ray nondetection |
| D. Xiang et al. (2024) | <−1876 | 70 | 15.70 | −5.12 | −18.72 | Prog. SED model |
| This paper | 1.51 | 25 | 14.57 | … | … | He I profile measurement |

**Note.** This table has been organized by listing the CSM radius in ascending order.
[a] Fiducial log($r$/cm) = 15 in order to calculate CSM density.

continuously encountered pre-shock CSM, from day 1.5 to day 3.5, suggesting a localization of mass loss between 1.1 and $3.7 \times 10^{14}$ ($v_{\rm shock}$/9500 km s$^{-1}$) cm. Observations of pre-SN activity constrain any luminous outburst to be brief ($\leqslant$200 days; Y. Dong et al. 2023) and dim (>–7 mag; C. L. Ransome et al. 2024), suggesting that no precursor events driven by radiation pressure occurred.

Mass-loss rates and wind velocities published prior to this work are listed in Table 3. The wind density at $r_{\rm CSM}$, the radius of CSM probed by the measurement, is listed and was calculated assuming an isotropic wind. The table is organized in order of ascending $r_{\rm CSM}$. A wide variety of wind densities and wind speeds are reported. This is attributable in part to different wavelength regimes, to methods that are sensitive to different radii of CSM, and to different assumptions of CSM velocities. Also important is that estimates of mass-loss density using progenitor SED modeling may have been derived from images older than the enhanced mass-loss episode. Hence, these estimates may only be sensitive to the RSG-like wind.

## 5. Conclusions

We have presented high-resolution echelle spectroscopy of SN 2023ixf beginning 1.51 days after explosion that monitors rapid changes in narrow emission lines associated with interaction between the SN and nearby CSM. The major findings and conclusions from our data set are as follows:

1. He I is present 1.51 days after explosion, but vanishes on day 2.62. This fleeting profile provides the most direct constraint on the pre-SN CSM speed. The speed of the slowest material in the CSM prior to the SN explosion was $25^{+0}_{-5} \pm 2$ km s$^{-1}$, as given by the day 1.51 He I emission line with uncertainties associated with an unknown amount of radiative acceleration and measurement of the line profile. This speed is within the normal range for a RSG wind (10–30 km s$^{-1}$; S. R. Goldman et al. 2017).
2. Lines across the spectrum and across epochs vary in width between 25 and 700 km s$^{-1}$ as radiation from the SN photosphere and shock–CSM interaction accelerates the CSM in the first 4 days following explosion.
3. We do not detect narrow P Cygni profiles in the pre-shock CSM at any early epoch ($t$ < 7 days), but we observe them in later spectra, likely arising from absorption by the post-shock CSM. Unlike the interaction between the SN and an isotropic CSM predicted by `HERACLES/CMFGEN`, the lack of a narrow P Cygni profile may suggest a asymmetric CSM.
4. The progenitor star underwent an enhanced mass-loss phase beginning $\gtrsim$4 yr prior to SN 2023ixf. By the time of explosion, the densest CSM extended out to $3.7 \times 10^{14}$ ($v_{\rm shock}$/9500 km s$^{-1}$) cm.
5. The r1w6b `HERACLES/CMFGEN` models with $\dot{M} = 0.01$ $M_\odot$ yr$^{-1}$ at high resolution (W. V. Jacobson-Galán et al. 2023) is consistent with the high-resolution echelle spectra presented here with respect to overall line shape, duration of narrow features, and time evolution. This mass-loss rate cannot be driven by typical RSG winds, which requires an energetic mechanism associated with terminal massive stellar evolution.

As demonstrated here, flash features encoding pristine information about pre-SN mass loss can change rapidly within the first 48 hr following core collapse. Additional spectroscopic observations of Type IIP SNe (SNe IIP) with high resolution are needed at these early epochs to study the full range of RSG pre-SN mass-loss velocities and strengths, which in turn can provide needed constraints on the nature of the pre-SN circumstellar environment (W. V. Jacobson-Galán et al. 2024b). Important contributions will be enabled by upcoming UV surveys capable of identifying SNe IIP during shock breakout. The UVEX and ULTRASAT missions (S. R. Kulkarni et al. 2021; Y. Shvartzvald et al. 2024) will regularly find SNe within a day of explosion and provide alerts enabling rapid follow-up opportunities. Finally, in order to make progress in understanding the distribution of CSM and the presence or absence of P Cygni profiles in narrow emission lines, further development of radiative transfer models that explore potential complex and asymmetric mass-loss environments is needed.

## Acknowledgments

We thank the referee for providing meaningful suggestions and comments that have significantly improved the manuscript's content and presentation. Kitt Peak National Observatory sits atop I'oligam Du'ag (Manzanita Shrub Mountain). Astronomers are honored to be permitted to conduct scientific research on the sacred mountain located in the homelands of the Schuk Toak District within the Toronto O'odham Nation. We honor their past, present, and future generations, who have lived here for time immemorial and will forever call this place home. We acknowledge the traditional homelands of the Indigenous People which Purdue University is built upon. We honor the Bodéwadmik (Potawatomi), Lenape (Delaware), Myaamia (Miami), and Shawnee People who are the original Indigenous caretakers.

We thank P. Duffell, W. Jacobson-Galán, A. Gal-Yam, A. Boestrom, and D. Hasenour for helpful guidance and discussion in the development of this work. Data presented here were obtained by the NEID spectrograph built by Pennsylvania State University and operated at the WIYN Observatory by NSF NOIRLab, under the NN-EXPLORE partnership of the National Aeronautics and Space Administration and the U.S. National Science Foundation. We thank the NEID Queue Observers and WIYN Observing Associates for their skillful execution of our NEID observations. Based on observations at Kitt Peak National Observatory, NSF's National Optical-Infrared Astronomy Research Laboratory (NOIRLab Prop. ID: 2021B-0455; PI: D. Milisavljevic), which is operated by the Association of Universities for Research in Astronomy (AURA) under a cooperative agreement with the National Science Foundation. D.M. acknowledges NSF support from grants PHY-2209451 and AST-2206532. D.J.H. gratefully acknowledges partial support through NASA Astrophysical Theory grant No. 80NSSC20K0524.

*Facilities:* NEID (C. Schwab et al. 2016), Supra Solem Observatory (https://www.suprasolemobs.com/).

*Software:* `astropy` (Astropy Collaboration et al. 2013), `matplotlib` (J. D. Hunter 2007), `SciPy` (P. Virtanen et al. 2020), `NumPy` (C. R. Harris et al. 2020), `pandas` (The pandas development team 2020).

## Appendix
## Functional Comparisons to the Spectra

This appendix contains line profiles for the emission lines and epochs listed in Table 2. Spectra are shown in Figures 11, 12, and 13. Overplotted in these figures are the Gaussians measured to the profiles with widths that correspond to the FWHM quoted in Table 2.

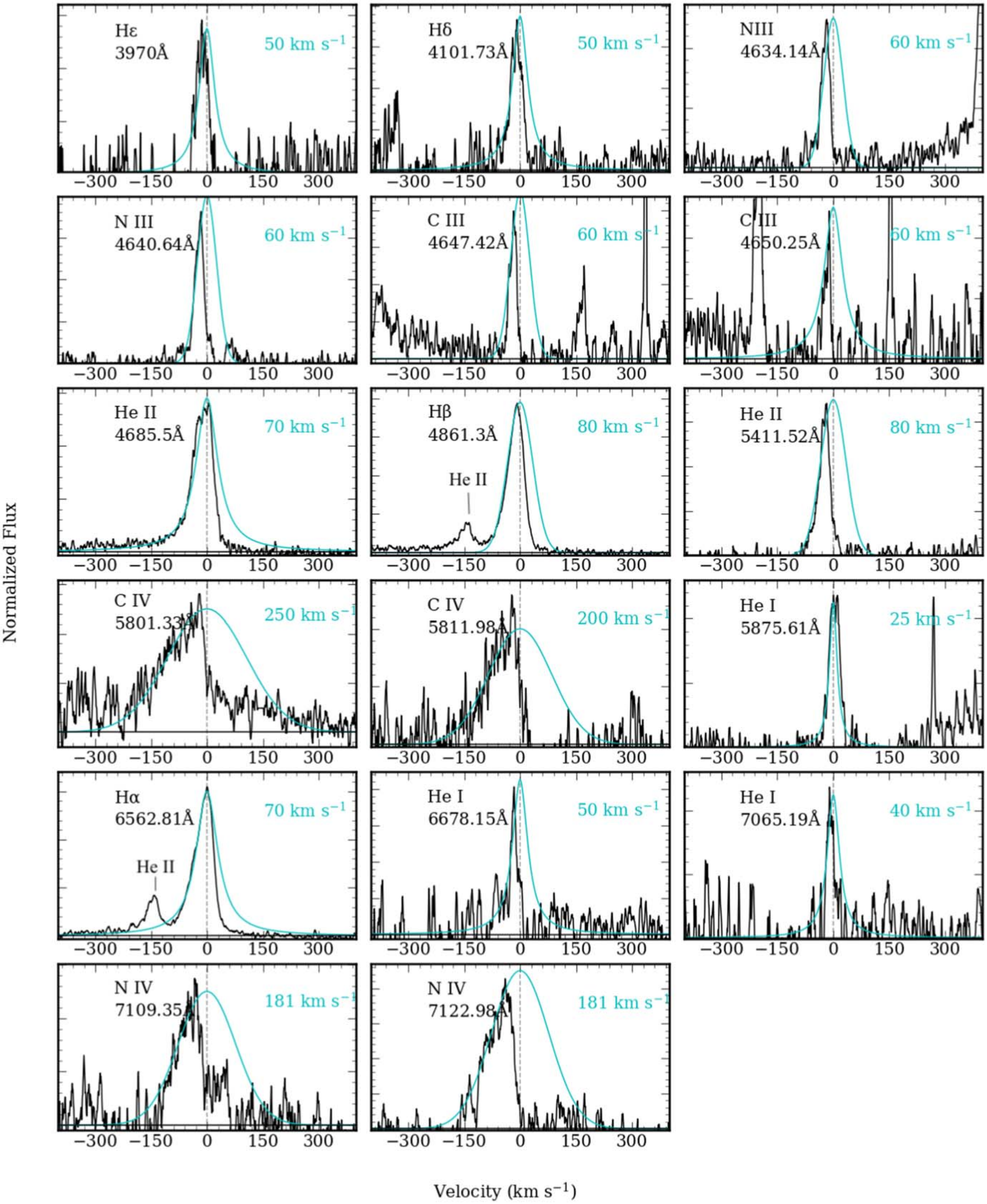

**Figure 11.** Various lines in the day 1.51 spectra. Overplotted are the functions used to measure the blue wing of each line. These velocities are listed in Table 2. The intermediate-width component of each line has been subtracted, leaving the narrow component. Spectra have been convolved with a 0.66 Å filter.

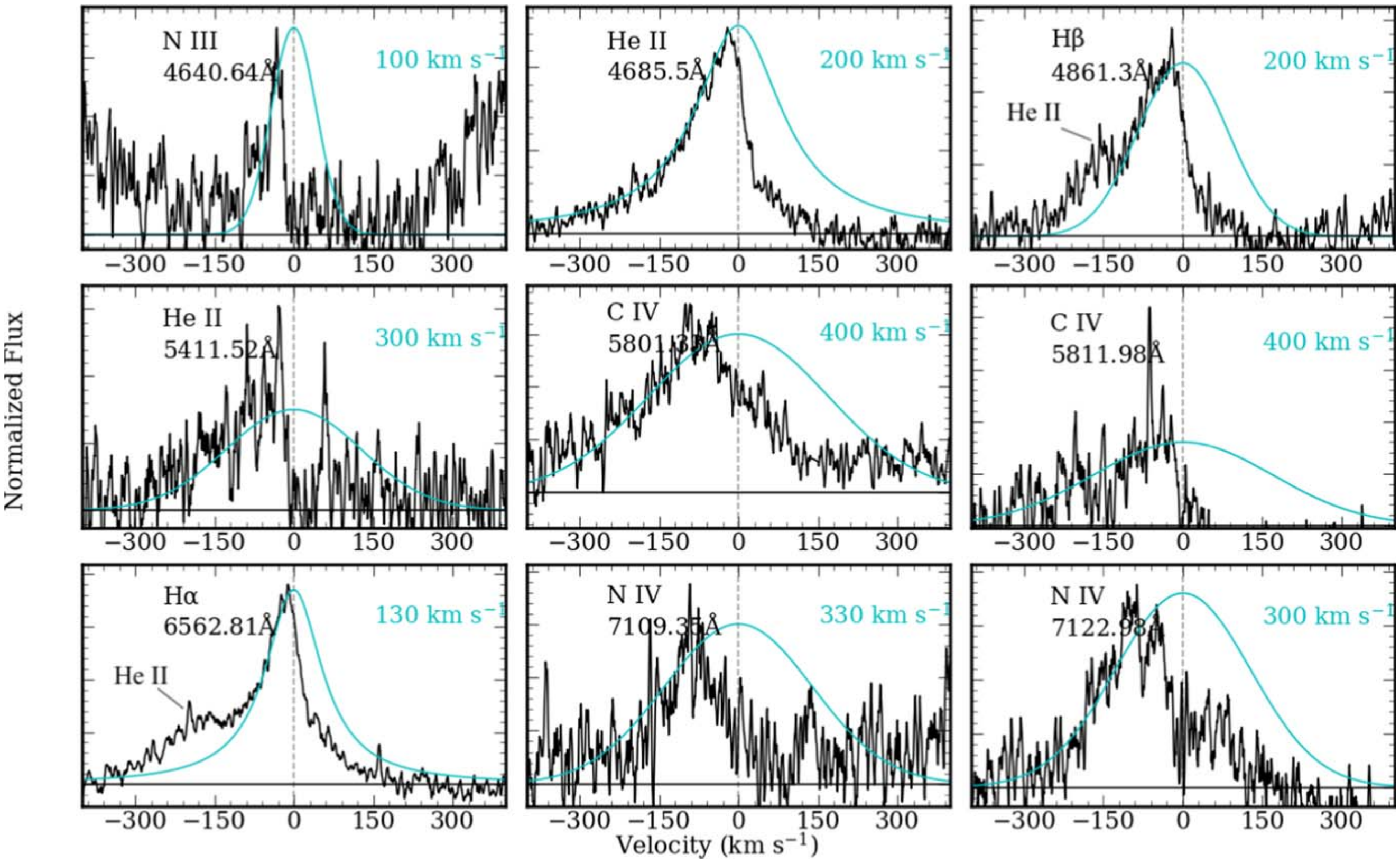

**Figure 12.** Same as Figure 11 but for day 2.62.

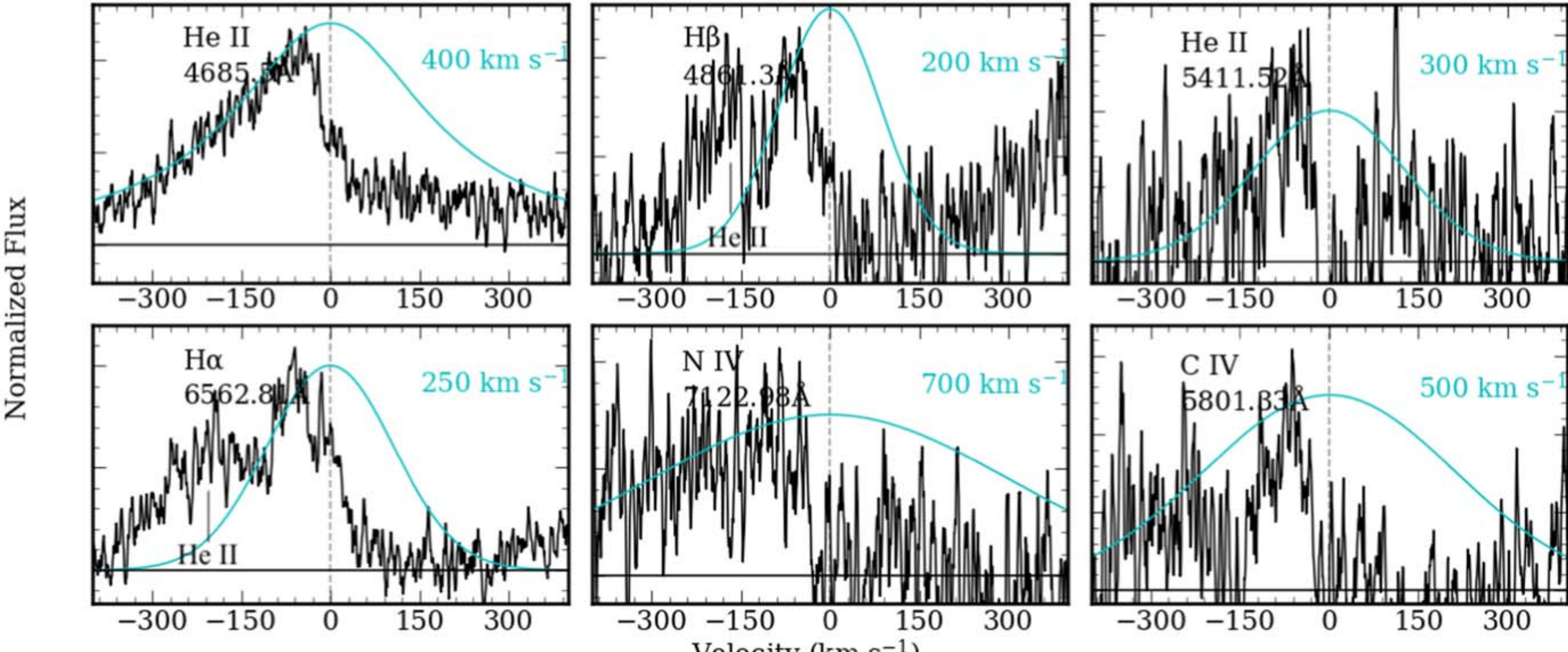

**Figure 13.** Same as Figure 11 but for day 3.47.

## ORCID iDs

Danielle Dickinson https://orcid.org/0000-0003-0913-4120
Dan Milisavljevic https://orcid.org/0000-0002-0763-3885
Braden Garretson https://orcid.org/0000-0001-6922-8319
Luc Dessart https://orcid.org/0000-0003-0599-8407
Raffaella Margutti https://orcid.org/0000-0003-4768-7586
Ryan Chornock https://orcid.org/0000-0002-7706-5668
Bhagya Subrayan https://orcid.org/0000-0001-8073-8731
D. John Hillier https://orcid.org/0000-0001-5094-8017
Dan Li https://orcid.org/0000-0001-7318-6318
Sarah E. Logsdon https://orcid.org/0000-0002-9632-9382
Jayadev Rajagopal https://orcid.org/0000-0002-2488-7123
Susan Ridgway https://orcid.org/0009-0008-5486-4433
Nathan Smith https://orcid.org/0000-0001-5510-2424